\documentclass[12pt,a4paper]{article}
\usepackage{times}
\usepackage{graphics}
\usepackage{graphicx}
\usepackage{float}
\restylefloat{figure}

\begin{document}
%\large
\date{ }

\begin{center}
{\Large  Constraints on strongly coupled chameleon fields from the experimental
test of the weak equivalence principle for the neutron.}

\vskip 0.7cm

Yu. N. Pokotilovski\footnote{e-mail: pokot@nf.jinr.ru}

\vskip 0.7cm
            Joint Institute for Nuclear Research\\
              141980 Dubna, Moscow region, Russia\\
\vskip 0.7cm

{\bf Abstract\\}

\begin{minipage}{130mm}

\vskip 0.7cm
 The chameleon scalar field is considered as a possible cause of accelerated
expansion of the Universe.
 The chameleon field induces an interaction potential between particle
and massive body.
 Previous experiments with falling cold neutrons intended to measure the
neutron coherent scattering lengths and verification of the weak equivalence
principle for the neutron are used to constrain the parameters characterizing
the strength of the scalar chameleon fields.

\end{minipage}
\end{center}
\vskip 0.3cm

PACS: quad 95.36.+x;\quad 04.80.cc; \quad 03.75.Be

\vskip 0.2cm


\vskip 0.6cm

 One of the most pressing mysteries in physics and cosmology is discovery of
the accelerated expansion of the Universe.
 The nature of this effect is not understood.
 Amongst several theoretical schemes proposed to explain this astronomical
observation one is a new cosmological scalar field of the quintessence type
\cite{Ratra} dominating the present day density of the Universe (recent reviews
are for example \cite{Pee,Cop}).

 Acting on cosmological distances the mass of this field should be very small --
of order of the Hubble constant: $\hbar H_{0}/c^{2}=10^{-33}\,eV/c^{2}$.

 The scalar fields appearing in modern string and supergravity theories should
couple to matter with gravitational strength.
 Direct coupling of light scalar fields to matter with a strength of
gravitation leads to large violation of the equivalence principle.
 But the experimental data yield very strict constraints on such a field
demanding their coupling to matter to be unnaturally small.

 The particular variant of the scalar field coupling to matter proposed in
\cite{Kho,Bra04,Gub,Upa06,Mota06,Mota07} has a form that in result of
self-interaction and interaction of the scalar field with matter the mass of
the scalar field depends on the local matter environment.

 In the proposed theory coupling of scalar field to matter is of the order as
demanded by string theory, but is very small on cosmological scales.
 In high matter density surrounding, according to the proposed field equations,
the mass of the field is increased, the interaction range is strongly decreased,
and the equivalence principle is not violated in laboratory experiments for the
search for the long range fifth force.
 The scalar field is confined inside the matter screening its existence to the
external world.

 The chameleon fields constructed in this way do not contradict to laboratory
tests of the equivalence principle and the fifth force experimental searches
even if these fields are strongly coupled to matter.
 In result of the screening effect the laboratory gravitational experiments
of Galileo-, E\"otv\"os- or Cavendish-type \cite{Fisch} performed with
macro-bodies at macroscopic distances are unable to set an upper limit on the
strength of the chameleon-matter coupling.
 At smaller distances ($10^{-7}-10^{-2}$) cm the new forces can be observed in
measurements of the Casimir force between closely placed macro-bodies
\cite{Rep}) or in the atomic force microscopy experiments.
 Casimir force measurements may evade to some degree the screening and probe
the interactions of the chameleon field at the micrometer range despite the
presence of the screening effect \cite{Mota07,Cas,Cas1}.

 It was shown in \cite{BraPi} that the chameleon interaction of elementary
particles with bulk matter should not be screened -- the chameleon induced
interaction potential of bulk matter with neutron can be in principle observed.
 It was proposed also in \cite{BraPi} to search for chameleon field through the
energy shift of ultracold neutrons in vicinity of reflecting horizontal mirror.
 From already performed experiments on observation of gravitational levels of
neutrons the constraints were obtained in \cite{BraPi} on parameters,
characterizing the force of chameleon-matter interaction.

 Chameleons can also couple to photons.
 In \cite{Ahl,Gies} it was shown that the chameleon-photon coupling leads to
the afterglow effect in a closed vacuum cavity in magnetic field.
 The continuing GammeV-CHASE \cite{CHASE,CHASE1} and ADMX \cite{ADMX}
experiments based on the proposal of \cite{Ahl,Gies} are intended to measure
(constrain) the coupling of chameleon scalar field to matter and photons.

 In the approach proposed here only chameleon-matter interaction is taken into
account not relying on existence of the chameleon-photon interaction.

 According to the chameleon scalar field theory
\cite{Kho,Bra04,Gub,Upa06,Mota06,Mota07} the chameleon effective potential is
%1
\begin{equation}
V_{eff}(\phi)=V(\phi)+e^{\beta\phi/M_{Pl}}\rho,
\end{equation}
where $V(\phi)$ is the scalar field potential:
%2
\begin{equation}
V(\phi)=\Lambda^{4}+\frac{\Lambda^{4+n}}{\phi^{n}},
\end{equation}
and $\rho$ is the local energy density of the environment.
 In these expressions $\Lambda=(\hbar^{3}c^{3}\rho_{d.e.})^{1/4}$=2.4 meV is
the dark energy scale, $\rho_{d.e.}\approx 0.7\times 10^{-8}$ erg/cm$^{3}$ is
the dark energy density.

 The chameleon interaction potential of a neutron with bulk matter
(in our consideration Earth's surface) was calculated in \cite{BraPi}:
%3
\begin{equation}
V(z)=\beta\frac{m}{M_{Pl}\lambda}\Bigl(\frac{2+n}{\sqrt{2}}\Bigr)^{2/(2+n)}
\Bigl(\frac{z}{\lambda}\Bigr)^{2/(2+n)}= \beta\cdot 0.9\cdot 10^{-21}\, eV
\Bigl(\frac{2+n}{\sqrt{2}}\Bigr)^{2/(2+n)} \Bigl(\frac{z}{\lambda}\Bigr)^{2/(2+n)},
\end{equation}
where $\lambda=\hbar c/\Lambda=82\,\mu m$.

 In obtaining constraints on strength of the chameleon field expressed by
the parameter $\beta$ we use the results of the experiments of Koester \cite{Koe}
who measured the critical height of reflection of free neutrons falling in
the Earth's gravitational field and reflected from horizontal liquid lead and 
bismuth mirrors.

 The acceleration of the free atom $g_{micro}$ in the Earth's gravitational
field has been measured by the Stanford group in 1999 with an accuracy
$3\times 10^{-9}$ \cite{Chu}.
 They also compared their result with the value of $g_{macro}$ obtained at the
same laboratory site using a Michelson interferometric gravimeter.
 It was found that the macroscopic object used in this measurement falls with
the same acceleration to within $7\times 10^{-9}$.
 From thus experimentally confirmed universality of free fall it follows that
the gravitational accelerations of free micro-particles (including neutron) and
of bulk matter are equal, and the weak principle of equivalence for
micro-objects established experimentally with precision better than $10^{-8}$.

 By the free fall in the Earth's gravitational field neutrons gain an energy
$mgh$, and if this energy is equal to the Fermi potential of the reflecting
horizontal mirror
%4
\begin{equation}
mgh=\frac{\hbar^{2}}{2m}4\pi Nb
\end{equation}
where $m$ is the neutron mass, $h$ is the fall's height, $N$ is the number
of nuclei in a unit volume of a mirror material, $b$ is the coherent
scattering length on a bound nucleus of the mirror, $h$ may be considered as a
critical height for the reflection.
 With measured with high precision $N$, $g$ and $h$ one obtains the neutron
scattering lengths.

 Koester compared neutron scattering lengths measured in the gravitational
diffractometer and those obtained by the neutron diffraction and scattering 
methods independent of gravity.
 The result of this comparison may be expressed by the factor $\gamma$
expressing the ratio of the neutron scattering lengths obtained by two methods:
$\gamma=1\pm 2.5\times 10^{-4}$.

 Schmiedmayer \cite{Sch} used all available data on scattering lengths and took
into account all systematic errors improving precision almost two times:
$\gamma=1\pm 1.7\times 10^{-4}$.

 The experiments of Koester \cite{Koe} may be interpreted as precision
measurement of the neutron potential energy above the Earth's surface, based on
an assumption that the neutron coherent scattering lengths do not depend on the
method of measurement and, therefore, on independent knowledge of the Fermi
potential of the mirror and of the Earth's gravitational acceleration for
microscopic (and macroscopic) bodies.

 As was mentioned for macroscopic bodies the chameleon shielding effect should
eliminate the effect of the scalar chameleon field on the acceleration of free
fall.
 But presence of additional chameleon-induced interaction should change the
potential energy of a neutron in vicinity of the Earth's surface.

 We can use the uncertainty of the Koester's measurements \cite{Koe}
and the Schmiedmayer's additional consideration \cite{Sch} to obtain the upper
limit of the effect of the chameleon field on a neutron's free fall
acceleration in vicinity of the Earth's surface:
%5
\begin{equation}
\Delta V(h)=\beta\cdot 0.9\cdot 10^{-21}\, eV
\Bigl(\frac{2+n}{\sqrt{2}}\Bigr)^{2/(2+n)}
\Bigl(\frac{h}{\lambda}\Bigr)^{2/(2+n)}\leq \Delta (gmh)
\end{equation}

 The constraints on parameter $\beta$ for different $n$ obtained from Eq. (6)
are shown in Fig. 1.
 The value of $h$ was taken 62 cm -- the critical height for bismuth.

 Existing constraints on the parameters $\beta$ and $n$ of the chameleon field
potential of Eq. (1) are not strong.
 From the atomic physics it follows \cite{at}, that $\beta\le 10^{14}$.
 More serious limits obtained from the experiments, demonstrating quantum
levels of neutrons reflecting from horizontal mirror in the Earth's
gravitational field, may be found in Fig. 1 of Ref. \cite{BraPi}.
 For example the allowed range of parameters for the strong coupling regime
$\beta\gg 1$ are: $50<\beta<5\times 10^{10}$ for n=1,
$10<\beta<2\times 10^{10}$ for n=2, and $\beta<10^{10}$ for $n>2$.

 It is seen that the result following from our work constraints the strength of
the chameleon field in the large coupling area of the theory parameters:
$\beta\leq 1.3\times 10^{7}$ for $n=1$,\, $\beta\leq 6\times 10^{7}$ for $n=2$
with weaker constraints for $n>2$ but still better than existing
constraints \cite{BraPi}.

%================================================================

\newpage
\begin{figure}
\begin{center}
\resizebox{13cm}{13cm}{\includegraphics[width=\columnwidth]{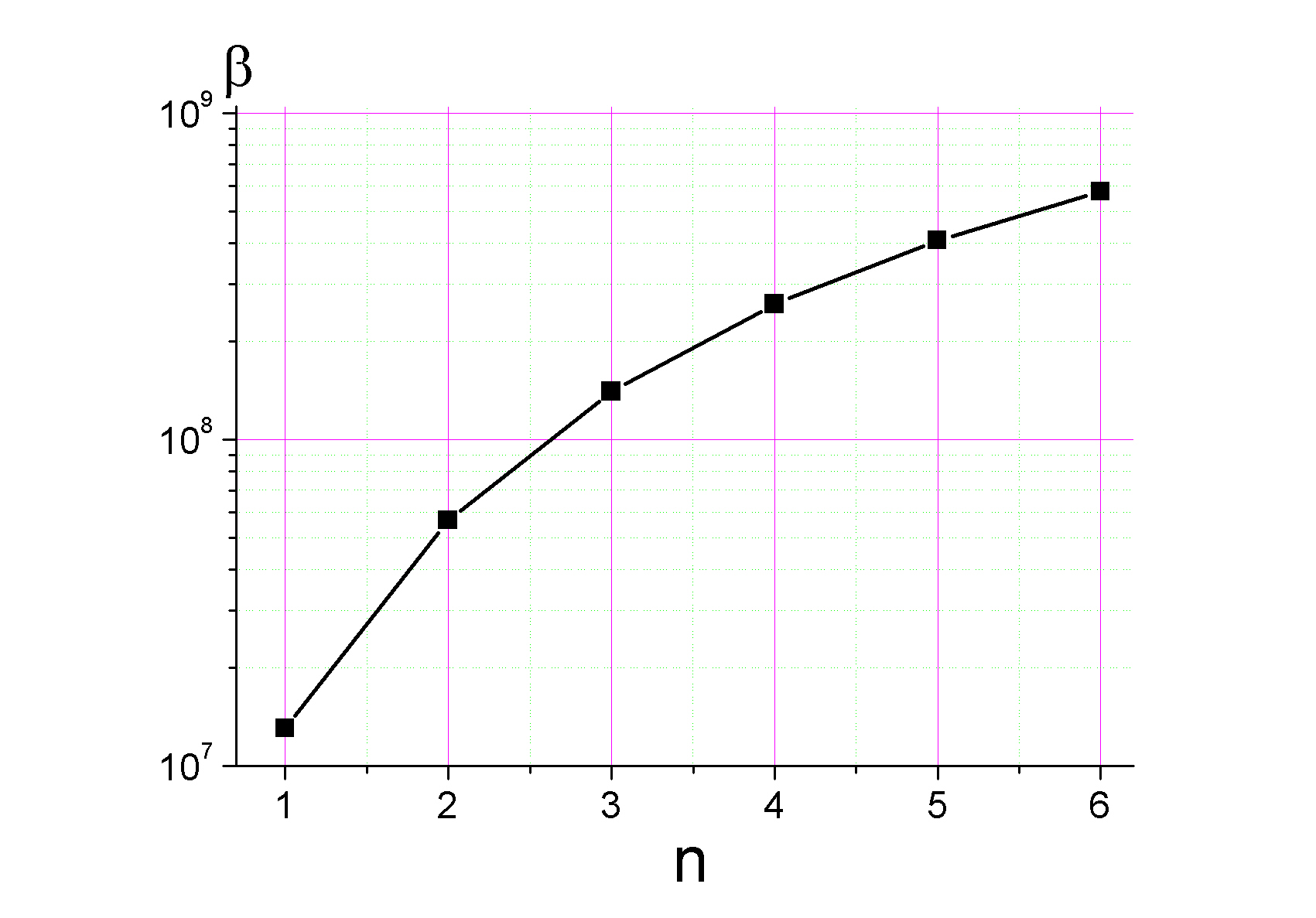}}
%\resizebox{15cm}{15cm}{\includegraphics[0cm, 0cm][15cm, 15cm]{fi1.JPG}}
\end{center}
\caption{Constraints on the parameter $\beta$ of the chameleon potential in 
dependence on $n$. Allowed region is below the curve.}
\end{figure}

\end{document}